\documentstyle[sprocl,epsfig]{article}

\def\be{\begin{equation}}
\def\ee{\end{equation}}
\def\bea{\begin{eqnarray}}
\def\eea{\end{eqnarray}}
\begin{document}

\title{SINGLE SPIN ASYMMETRIES IN $p^\uparrow p$, $\bar{p}^\uparrow p$\\
INCLUSIVE PROCESSES}

\author{M. Anselmino$^a$, M. Boglione$^b$, F. Murgia$^c$}

\vspace{3pt}

\address{
\baselineskip=7pt
$^a$Dipartimento di Fisica Teorica, Universit\`a di Torino and\\
 INFN, Sezione di Torino, Via P. Giuria 1, 10125 Torino, Italy\\
$^b$Department of Physics and Astronomy, Vrije Universiteit Amsterdam\\
 De Boelelaan 1081, 1081 HV Amsterdam, The Netherlands\\
\vspace{2pt}
$^c$Dipartimento di Fisica, Universit\`a di Cagliari and\\
 INFN, Sezione di Cagliari, C.P. n. 170, I-09042 Monserrato (CA), Italy}


\vspace{-3pt}

\maketitle\abstracts{Single spin asymmetries in $p^\uparrow
(\bar p^\uparrow) \> p \to \pi\,X$ have been measured
to be large. They can be understood within perturbative
QCD by introducing new $\mbox{\boldmath$k$}_\perp$ and spin dependences
in fragmentation and/or distribution functions. We discuss first how these
effects in distribution functions alone can account for existing data
and predict single spin asymmetries in other processes; then we see how
similar effects in fragmentation functions alone can fit the existing data;
finally we devise strategies to discriminate between the two contributions.}

\vspace{-5pt}

We have recently presented a formalism to evaluate
single spin asymmetries (SSA) in inclusive particle production
at high energy and moderately large $p_T$ \cite{abm,am2}.
This formalism is based on the parton model approach
and on factorization theorems, generalized to polarized
reactions and to the inclusion of transverse momentum effects in the
partonic distributions of the initial, polarized proton (the so-called
``Sivers'' effect). This requires
the introduction of a new, soft, nonperturbative distribution function
(non-diagonal in the helicity basis), $\Delta_N\tilde{f}$. It has been
shown \cite{abm} how recent experimental results on SSA for pions in
the fragmentation region \cite{e704} can be well reproduced
by opportunely parametrizing $\Delta_N\tilde{f}$. This fit to
$\Delta_N\tilde{f}$ has been in turn used to give
predictions for SSA in several other interesting processes \cite{am2}.

In this contribution we present a complete generalization of our formalism,
with the inclusion of partonic transverse momentum effects {\it both} in the
partonic distributions of the initial polarized hadron (Sivers effect)
{\it and} in the fragmentation process (the so-called ``Collins'' effect).
%We will then discuss if and how the Collins effect
%alone is able to reproduce the pion SSA \cite{e704} (as already
%mentioned before, the Sivers effect has been
%discussed at length in Ref.s \cite{abm,am2}).
%We will also discuss a possible strategy to discriminate
%between these two competing effects, which in the case of pion
%(and, in general, hadron) production may be present
%at the same time.
Due to space limitations, we only give the final result
of our formalism, in the case of $p^\uparrow p\to\pi\,X$ processes.
A more detailed discussion will be presented elsewhere \cite{abmm}
(see also Ref.s 1,2).
At leading order in transverse momentum effects, we have:

\vspace{-8pt}

\begin{eqnarray}
2\> \frac{E_\pi \, d\sigma^{p^\uparrow p\to \pi\,X}}{d\mbox{\boldmath$p$}_\pi}
\> A_N(p^\uparrow p\to \pi + X) \simeq \qquad\qquad\qquad
\qquad\qquad\qquad\qquad\nonumber\\
\sum_{abcd}\int\frac{dx_adx_b}{\pi z}\,
d^2\mbox{\boldmath$k$}_\perp\,
%\Bigl[\tilde{f}_{a/p^\uparrow}(x_a,
%\mbox{\boldmath$k$}_\perp)-\tilde{f}_{a/p^\uparrow}(x_a,
%-\mbox{\boldmath$k$}_\perp)\Bigr]
\Delta_N\tilde{f}_{a/p^\uparrow}(x_a,\mbox{\boldmath$k$}_\perp)
\,f_{b/p}(x_b)\,
\frac{d\hat{\sigma}^{ab\to cd}}{d\hat{t}}(x_a,x_b,
\mbox{\boldmath$k$}_\perp)\,D_{\pi/c}(z) \nonumber\\
+ \sum_{abcd}\int\frac{dx_adx_b}{\pi z}\,
d^2\mbox{\boldmath$k$}'_\perp\,\Delta_T f_{a/p}(x_a)\,f_{b/p}(x_b)
\,\Delta_{NN}\hat{\sigma}(x_a,x_b,\mbox{\boldmath$k$}'_\perp)
\,\Delta_T\tilde{D}_{\pi/c}(z,\mbox{\boldmath$k$}'_\perp)\nonumber\\
\label{an}
\end{eqnarray}

\vspace{-8pt}

The second line in Eq.~(\ref{an}) accounts for effects in the distribution
functions (Sivers effect). It contains a new, soft distribution,
$\Delta_N\tilde{f}_{a/p^\uparrow}(x_a,\mbox{\boldmath$k$}_\perp) \equiv 
\tilde{f}_{a/p^\uparrow}(x_a,
\mbox{\boldmath$k$}_\perp)-\tilde{f}_{a/p^\uparrow}(x_a,
-\mbox{\boldmath$k$}_\perp)$, which depends on the transverse momentum
$\mbox{\boldmath$k$}_\perp$ of parton $a$ inside the polarized proton
$p^\uparrow$. Notice that $\Delta_N\tilde{f}_{a/p^\uparrow}$ is an odd
function of $\mbox{\boldmath$k$}_\perp$,
so that the Sivers effect vanishes if we let
$\mbox{\boldmath$k$}_\perp\to 0$: we have to take
into account $\mbox{\boldmath$k$}_\perp$ effects
also in the elementary partonic cross section, which makes the total
effect a twist-three contribution. At leading order in transverse
momenta, we can neglect
in this term $\mbox{\boldmath$k$}'_\perp$ effects in the fragmentation.
Analogously, the third line in Eq.~(\ref{an}) refers to the
Collins effect in the fragmentation process. Here we can in turn
neglect $\mbox{\boldmath$k$}_\perp$ effects in the distribution functions.
$\Delta_T\tilde{D}_{\pi/c}(z,\mbox{\boldmath$k$}'_\perp) \equiv
\tilde{D}_{\pi/c^\uparrow}(z,\mbox{\boldmath$k$}'_\perp)-
\tilde{D}_{\pi/c^\uparrow}(z,-\mbox{\boldmath$k$}'_\perp)$ plays the same
role of $\Delta_N\tilde{f}_{a/p^\uparrow}$, this time in the $c^\uparrow 
\to \pi + X$ fragmentation process. Due to the initial polarized proton,
the Collins contribution is more complicated than the Sivers' one.
In particular, it involves the so-called transversity distribution
$\Delta_Tf_{a/p}$ (or $h_1$) for parton $a$ inside the transversely polarized 
proton (also unknown) and $\Delta_{NN}\hat{\sigma} \equiv
d\hat\sigma^{a^\uparrow b\to c^\uparrow d}/d\hat t-
d\hat\sigma^{a^\uparrow b\to c^\downarrow d}/d\hat t$.

A simplified version of Eq.~(\ref{an}) is used for
practical calculations \cite{abm,am2}.
We consider only valence parton contributions to
$\Delta_N\tilde{f}$ and $\Delta_TD$, which is quite resonable, in
particular in the fragmentation region where
sizeable SSA are observed \cite{e704}.
Moreover, the full dependence on transverse momentum (both for
$\mbox{\boldmath$k$}_\perp$ and $\mbox{\boldmath$k$}'_\perp$)
is simplified by introducing a fixed, average transverse momentum,
which sets the relevant physical scale for the overall effect.
The explicit $x_a$ ($z$) dependence in $\mbox{\boldmath$k$}_\perp$
($\mbox{\boldmath$k$}'_\perp$) can be deduced from theoretical information
or extrapolated from experimental data.
The residual $x_a$ ($z$) dependence in $\Delta_N\tilde{f}$
($\Delta_TD$) is in turn parametrized by simple functions
of the form $Nu^a(1-u)^b$, where $u=x_a$ or $z$.

\begin{figure}[t]
\vspace{-1.0truecm}
\begin{center}
\epsfig{figure=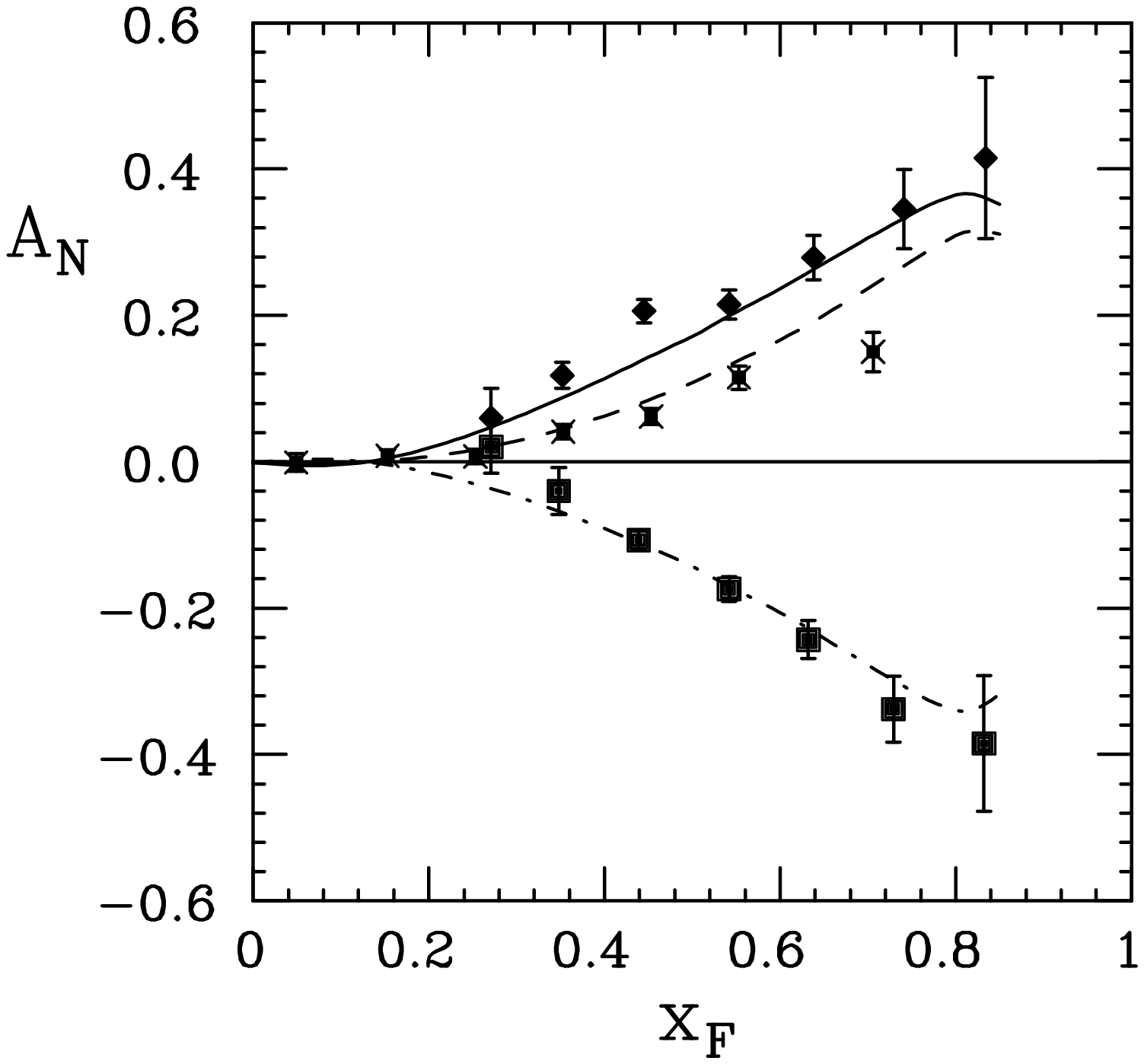,bbllx=50pt,bblly=200pt,bburx=530pt,%
bbury=650pt,height=7cm}
 \begin{minipage}[c]{10cm}
\baselineskip=3pt
 {\footnotesize {\bf Fig. 1:}
Fit to experimental data on pion single spin asymmetry with the Collins
contribution; the upper, middle, and lower sets of data and curves refer
respectively to $\pi^+$, $\pi^0$, and $\pi^-$.} 
 \end{minipage}
 \end{center}
\vspace{-13pt}
\end{figure}

It has been shown \cite{abm,am2} that Sivers effect alone is able
to reproduce the pion SSA with quite a reasonable parametrization
for $\Delta_N\tilde{f}$. In Fig. 1 we show preliminary results
indicating that a comparatively good fit to the data can be obtained
when only Collins effect is active.

It is difficult, however, with the presently limited amount of
experimental information, to understand if and to what extent one
of the two contributions is more effective than the other in a given
process. To this end, more experimental data are clearly required
on different processes. A possible strategy which can help in
discriminating between Sivers and Collins effect is the following:
{\it i)} One can first consider the process $p^\uparrow p\to\gamma\,X$.
In this case there is no Collins contribution at all, and
the process gives information on $\Delta_N\tilde{f}$.
Predictions for this process have been given
(with $\Delta_N\tilde{f}$ coming from the fit to pion data, when only
Sivers effect is active) \cite{am2}.
{\it ii)}
As a second step, one can consider meson production in semi-inclusive
polarized DIS, {\it e.g.} $\ell p^\uparrow\to\ell'\pi\,X$. In this case
Sivers effect is ruled out by necessary initial state interactions, and one 
can get unambiguous information 
regarding Collins contribution and $\Delta_TD(z)$.
Work in this direction is in progress.
{\it iii)} Finally, one can again consider meson production,
{\it e.g.}, $p^\uparrow p\to\pi\,X$, etc., and check if and to what extent a
simultaneous fit of the three processes is possible,
with physically reasonable parametrizations of
$\Delta_N\tilde{f}$ and $\Delta_TD(z)$.

Let us stress that our formalism is the only one that
consistently accounts for two of the possible mechanisms suggested as
explanation for sizeable SSA in the fragmentation region.
A combined theoretical and experimental analysis of several processes
involving different final particles should allow to discriminate between
Sivers and Collins effect and help to test our model and other alternative
proposed models.

%\section*{Acknowledgments}


\vspace{-9pt}

\section*{References}
\begin{thebibliography}{99}

\vspace{-4pt}

\small

\bibitem{abm}  M.~Anselmino, M.~Boglione, and F.~Murgia,
               Phys. Lett. {\bf B362}, 164 (1995) and references therein.
\bibitem{am2}  M.~Anselmino, F.~Murgia, e-Print Archive: hep-ph/9808426.
\bibitem{e704} D.L. Adams {\it et al.}, Phys. Lett. {\bf B264}, 462 (1991).
\bibitem{abmm} M.~Anselmino, M.~Boglione, P. Mulders, and F.~Murgia,
               in preparation.

\end{thebibliography}
\end{document}